\documentclass[twocolumn,showpacs,prl]{revtex4}

\usepackage{amsmath}
\usepackage{amsfonts}            
\usepackage{amssymb}

\usepackage{mathrsfs}

\usepackage{eufrak}

\usepackage{verbatim} 

\usepackage{graphicx}
\usepackage{epstopdf}

\usepackage{dcolumn}
\usepackage{bm}

\begin{document}

\title{Probing weakly-bound molecules with nonresonant light}

\author{Mikhail Lemeshko}

\author{Bretislav Friedrich}

\affiliation{%
Fritz-Haber-Institut der Max-Planck-Gesellschaft, Faradayweg 4-6, D-14195 Berlin, Germany
}%

\date{\today}

\begin{abstract}

We show that weakly-bound molecules can be probed by ``shaking" in a pulsed nonresonant laser field. The field introduces a centrifugal term which expels the highest vibrational level from the potential that binds it. Our numerical simulations applied to the Rb$_2$ and KRb Feshbach molecules indicate that shaking by feasible laser pulses can be used to accurately recover the square of the vibrational wavefunction and, by inversion, also the long-range part of the molecular potential.

\end{abstract}

\maketitle


Weakly-bound species, such as haloes~\cite{JensenHalo04}, spend most of their vibrational period in the classically forbidden region of the electronic potential. The accuracy required to treat such a potential stretches current computational methods to their limits. Here we present a versatile technique for determining the interatomic potential from the dependence of the dissociation probability of the weakly-bound molecule on the intensity of a pulsed nonresonant laser field. We also show that a cw laser field can be used to control the molecules' size. 

Since diatomic molecules formed by photo- or Feshbach-association of ultracold atoms often emerge in the highest vibrational level of a ground electronic state~\cite{GrimmChapter09}--\cite{KohlerRMP06}, they represent prototypical quantum halo systems.  Among other prominent examples of haloes are the recently observed Efimov trimer~\cite{KraemerNature06} and tetramer~\cite{FerlainoPRL09}, the latter produced in collisions involving halo dimers \cite{KnoopNature09}. 

The technique relies on the ability to impart a value of angular momentum to the weakly-bound molecule such that the centrifugal term concomitant with it expels the molecule's vibrational level from the potential and thus causes the molecule to dissociate. Unlike in the case of rotational predissociation, the value of the imparted angular momentum is tunable, derived from the anisotropic polarizability interaction of the molecule with a nonresonant laser field. This interaction creates directional states in which the molecular axis librates (shakes) about the field vector \cite{FriHerPRL95}. The laser intensity needed to impart a preordained value of the angular momentum varies characteristically with the internuclear distance. It is this characteristic dependence that can be used to map out the probability density of the vibrational state from which the molecule was forced to dissociate. A highly accurate long-range molecular potential can be then obtained by inverting the vibrational probability density. This route to an accurate potential, independent of spectroscopy or scattering, complements what can be learned from either. We illustrate the technique's machinery by examining Feshbach molecules of acute interest, Rb$_2$ and KRb. 

Although the technique is applicable to weakly-bound molecules in any rotational state, we limit our considerations here to the case of rotationless (diatomic) molecules, i.e., species in vibrational levels that can only support the ground rotational state. In addition, we consider the molecule to be in a $^1\Sigma $ electronic state, with a $V(r\rightarrow \infty)=-C_6r^{-6}$ asymptotic potential. In order for a level to be rotationless, its binding energy, $E_b$, must fulfill the inequality $|E_b| <  d_6 \hbar^3 m^{-3/2}C_6^{-1/2}$, where $m$ is the reduced mass and $d_6 \approx 1.6$ is a dimensionless parameter which depends solely on the exponent of the power law potential and can be evaluated analytically~\cite{LemFri_BriefReport}. The centrifugal term in the effective potential
\begin{equation}
	\label{EffPot}
	U(r) = V(r) + \frac{\langle \mathbf{J}^2 \rangle \hbar^2}{2 m r^2}
\end{equation}
then leads to dissociation for any higher rotational states. Here $V(r) $ is the electronic potential and
$\left < \mathbf{J}^2 \right > = \left< \psi (t) \right | \mathbf{J}^2 \left |  \psi (t) \right >$ is the expectation value of the square of the imparted angular momentum in a state $\left |  \psi (t) \right >$ created by the anisotropic polarizability interaction~\cite{FriHerPRL95}. The critical value of the angular momentum, $J^{\ast}$, corresponding to $\langle \mathbf{J^{\ast}}^2 \rangle=J^{\ast}(J^{\ast}+1)$, for which the effective potential can no longer support a given vibrational level can be much smaller than unity. For instance, for the highest vibrational state, $v=123$, of $^{85}$Rb$_2 (^1\Sigma^+_g)$, we find $J^{\ast} =0.22$. Such an angular momentum can be imparted to the molecule by a feasible laser field. As noted earlier \cite{frigupher}, for $^{4}$He$_2 (^1\Sigma^+_g,v=0)$, the critical angular momentum $J^{\ast}$ is $0.0337$. 

We obtain the wavefunction $\left |  \psi (t) \right >$ by solving the time-dependent  Schr\"{o}dinger equation for a dimensionless Hamiltonian
\begin{equation}
\frac{H(t)}{B}= \mathbf{J}^{2} - g(t) \left[\Delta \omega \cos ^{2}\theta +\omega _{\bot
} \right]  \label{ham1}
\end{equation}
Here $g(t)$ is the time profile of a plane-polarized laser pulse of peak intensity $I$, $\Delta \omega \equiv \omega _{||}-\omega _{\bot }$, and $\omega _{||,\bot } \equiv 2\pi \alpha_{||,\bot }I/(Bc)$, with $\alpha _{||}(r)$ and $\alpha _{\bot }(r)$ the polarizability components parallel and perpendicular to the molecular axis, and $B=B(r) \equiv \hbar^2 / (2mr^2)$.
Because of the azimuthal symmetry about the field vector, the induced dipole potential involves just the polar angle $\theta$ between the molecular axis and the polarization plane of the laser pulse. We assume the oscillation frequency 
of the laser field to be far removed from any molecular resonance and much higher than the reciprocal of the pulse duration; as a result, the time dependence of the radiative field 
is reduced to that of the time profile \cite{Ortigoso99}. For homonuclear molecules, the dependence of the polarizability anisotropy, $\Delta \alpha (r)\equiv \alpha _{||}(r)-\alpha _{\bot }(r)$, on the internuclear distance is well captured at large $r$ by Silberstein's expansion, $\Delta \alpha (r)=6\alpha^2_0 r^{-3}+6\alpha_0^3 r^{-6}+...$, where $\alpha_0$ is the atomic polarizability~\cite{Silber}.
\begin{figure}
\includegraphics[width=8.5cm]{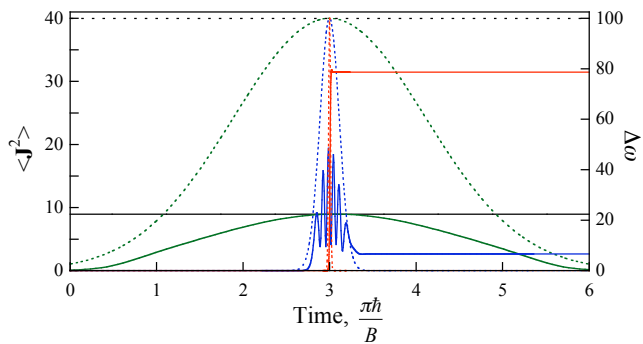}
\caption{\label{fig:J2} Temporal dependence of the angular momentum imparted to a molecule by laser pulses of duration $\tau=2.5~\pi \hbar/B$ (green solid line), $\tau=0.25~\pi \hbar/B$ (blue solid line), and $\tau=0.025~\pi \hbar/B$ (red solid line). The corresponding time profiles of the laser pulses, centered at time $3\pi \hbar/B$, are shown by dotted lines. Also indicated is the angular momentum due to a cw laser field (black solid line). Note that $\Delta \omega=100$ at the peak of the pulsed fields and for the cw field. See text.}
\end{figure}
The solutions of the reduced time-dependent Schr\"{o}dinger equation corresponding to Hamiltonian (\ref{ham1}), 
\begin{equation}
i\frac{\hbar }{B}\frac{\partial \psi (t)}{\partial t}=\frac{H(t)}{B}\psi (t),
\label{schr1}
\end{equation}
can be expanded in a series of field-free
rotor wave functions $|JM\rangle \equiv Y_{JM}$ (pertaining to eigenenergies 
$E_{J}$) 
\begin{equation}
\psi (\Delta \omega (t))=\sum_{J}c_{J}(\Delta \omega (t))|J,M\rangle \exp
\left( -\frac{iE_{J}t}{\hbar }\right)
\end{equation}
whose time-dependent coefficients, $c_{J}\left( \Delta \omega \left(
t\right) \right) $, solely determine the solutions at given initial
conditions (in the interaction representation). Thus the \emph{hybridization} coefficients $c_{J}$ can be
found from the differential equations 
\begin{multline}
i\frac{\hbar }{B}\stackrel{.}{c}_{J}(t)=-\sum_{J^{\prime }}c_{J^{\prime
}}(t)\left\langle J,M\right| \Delta \omega \cos ^{2}\theta +\omega _{\bot
}\left| J^{\prime },M\right\rangle \\ 
\times \exp \left[ -\frac{i\left( E_{J^{\prime
}}-E_{J}\right) t}{\hbar }\right] g(t)
\label{diff1}
\end{multline}
where we consider the pulse shape function to be a Gaussian, $g(t)=\exp \left[-4\ln(2)t^{2}/\tau^{2} \right]$, characterized by a full width at half maximum, $\tau$, the ``pulse duration.''  By taking into account the non-vanishing matrix elements $\left\langle
J,M\right| \cos ^{2}\theta \left| J^{\prime },M\right\rangle $, Eq. (\ref{diff1}) reduces to a tridiagonal form and can be
numerically solved, at any particular value of $r$, by standard methods~\cite{Ortigoso99}. We note that the expectation value of $B$ in a given vibrational state, $\phi_v \equiv | v \rangle$, is the rotational constant pertaining to that state, $B_v\equiv\langle v|B|v \rangle=\frac{\hbar^2}{2m}\langle v|r^{-2}|v \rangle$. Similarly, the expectation value $\Delta \alpha_v \equiv \langle v|\Delta \alpha|v \rangle$ is the polarizability anisotropy of the molecule in the vibrational state $|v\rangle$. 

Figure~\ref{fig:J2} shows the temporal dependence of the angular momentum imparted to a molecule at a time $\frac{3\pi \hbar}{B}$ for different values of the pulse duration $\tau$. One can see that  whereas for $\tau > 2.5~\pi \hbar/B$, the angular momentum of the molecule is enhanced only during the time the laser pulse is on (adiabatic limit), for $\tau <0.25 \pi~\hbar/B$ the molecule acquires a part of the transferred angular momentum for good (nonadiabatic limit). At $\tau <0.025~\pi \hbar/B$, \emph{most} of the angular momentum is imparted for good. Note that the amplitude of the laser pulse, $\Delta \omega =100$, is the same for all three pulses. The same value of $\Delta \omega$ was also used for the cw, stationary case, with $g(t)=1$, shown by the horizontal line.

Thus, in the nonadiabatic regime, characterized by $\tau \ll \pi \hbar/B$, a laser pulse can endow a molecule, on a timescale $\tau$, with a preordained value of angular momentum which the molecule will keep after the pulse has passed. For pulses shorter than the vibrational period, $\tau<\tau_v$, the interaction with the laser field will be well within the nonadiabatic regime. Moreover, in this case, the imparted angular momentum will also depend on the value of the internuclear distance, $r=r_p$, which the molecule had at the moment when the laser pulse struck. For disparate time scales  such that $\tau \ll \tau_v \ll \tau_r$, the dimer is transferred from the $V(r)$ to the $U(r)$ potential essentially instantaneously.

The laser intensity needed to impart a given value of angular momentum at a given internuclear distance can be found by solving Eq. (\ref{schr1}). For a  critical value $\langle \mathbf{J^{\ast}}^2 \rangle$ of the angular momentum, Eq. (\ref{schr1}) yields a laser intensity $I$ which corresponds to a critical internuclear distance $r^{\ast}$ such that when struck at $r_p = r^\ast$, the molecule will dissociate.  It is shown below that $I(r^{\ast})$ increases essentially monotonously with $r^{\ast}$. As a result, for a given intensity $I(r^{\ast})$, the molecule will receive a critical value of the angular momentum and dissociates if the pulse strikes at any distance $r_p \le r^\ast$. On the other hand, for $r_p > r^\ast$, the imparted angular momentum is subcritical and the molecule will remain intact. Hence the dissociation probability coincides with the probability of finding the molecule with an internuclear distance less than $r^\ast$,

\begin{equation}
	\label{DissProbGeneral}
	F(r^\ast) = \int_0^{r^\ast} \left | \phi_v(r) \right |^2 dr.
\end{equation}
The dependence, $F=F(I)$, of the dissociation probability $F$ on the intensity $I$ of the laser pulses is experimentally observable. By making use of the computed $I=I(r^{\ast})$ dependence, it can be transformed into $F(r^{\ast})$, whose derivative with respect to $r^{\ast}$ yields the vibrational probability density,  $ \left| \phi_v(r) \right |^2 $. Note that while $F(r^\ast)$ depends on the molecular  potential, $I(r^\ast)$ does not, as it reflects just the molecular moment of inertia and polarizability. In the short-pulse limit, $I(r^{\ast})$ has a universal asymptotic dependence on $r^{\ast}$,  $I(r^{\ast}\rightarrow \infty)\propto (r^{\ast})^3$, which can be derived analytically from the closed-form solution of Eq. (\ref{schr1}) in that limit, given in ref. \cite{caifriCCCC}. The polarizability anisotropy $\Delta \alpha(r)$ at large $r$ is accurately captured by Silberstein's formula~\cite{Silber}. Alternatively, the need to know $\Delta \alpha(r)$ can be eliminated by measuring $F(I)$ for two weakly-bound vibrational levels.

\begin{figure}
\includegraphics[width=8.5cm]{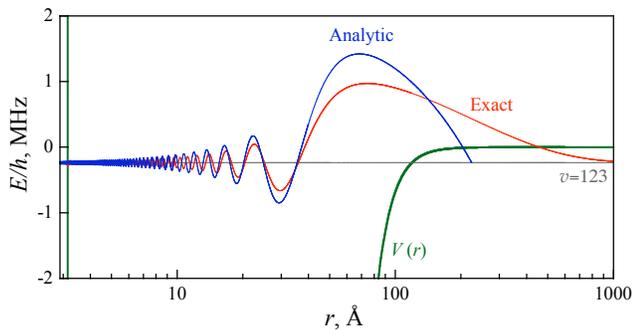}
\caption{\label{fig:Pots_WFs} The ``exact" single potential energy curve $V(r)$ of $^{85}$Rb$_2$ (green solid curve) and the wavefunctions of the highest vibrational state, $v=123$, obtained from the ``exact" potential (red solid line) and from the analytic model (blue solid line). The binding energy of the $v=123$ level is $E_b=-237$~kHz. See text.}
\end{figure}

We now illustrate the outlined procedure by evaluating the dissociation probability of the  $^{85}$Rb$_2$ molecule in its highest vibrational state. Based on the molecular constants and  a combination of the potential energy curve of ref.~\cite{Seto00} with the long-range exchange and dispersion terms of ref.~\cite{vanKempen02}, a single potential curve for the ground $^1\Sigma_g^+$ electronic state was found, with a dissociation energy $D_e=3993.53$~cm$^{-1}$ and an equilibrium distance $r_e = 4.21$~\AA.  An extrapolation of this ``exact" potential suggests that the highest bound state has $v=123$. We corroborated this result by numerically solving the Schr\"odinger equation for this potential, which also yielded the binding energy and the wavefunction of the $v=123$ state, see Figure~\ref{fig:Pots_WFs}. The $v=123$ vibrational state is no longer bound if the squared angular momentum is greater than $\langle \mathbf{J^{\ast}}^2 \rangle =0.27$. We found from Eq. (\ref{diff1}) in the limit $t \rightarrow \infty$ and for a dependence of the molecular polarizability on the internuclear distance taken from ref.~\cite{Deiglmayr08} that
such an angular momentum is imparted to the $^{85}$Rb$_2$ molecule by a cw laser field of intensity $I=6.58 \times 10^8$~W/cm$^2$.

\begin{figure}
\includegraphics[width=8.5cm]{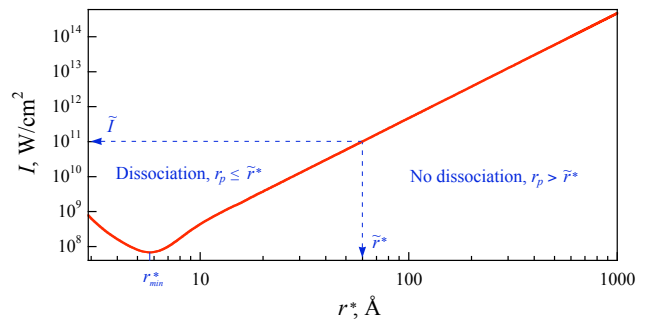}
\caption{\label{fig:Rb2_I_rp} Dependence of the intensity of 50~ps laser pulses needed for imparting a critical value of angular momentum to the Rb$_2(^1\Sigma^+,v=123)$ molecule on the critical internuclear distance $r^{\ast}$. See text.}
\end{figure}

In order to map out the wavefunction squared of the highest vibrational state, the molecule must be shaken by a pulsed laser field whose pulse duration is much shorter than the vibrational period. Since the vibrational period of halo molecules is often in excess of nanoseconds, the vibrational motion can be adequately parsed by 50 ps pulses, easily available in the laboratory \cite{ShortLasers}. The use of such pulses simultaneously ensures that the polarizability interaction will be well within the nonadiabatic regime, as $\tau \ll \tau_v \ll \frac{\pi \hbar}{B_v}$. However, since $\tau_v=0.67$ $\mu$s for the $v=123$ state of $^{85}$Rb$_2$, even ns pulses would suffice to probe it.

The calculated dependence of the intensity $I$ of 50 ps laser pulses needed to dissociate $^{85}$Rb$_2(v=123)$ on the critical internuclear distance $r^{\ast}$  is presented in Fig.~\ref{fig:Rb2_I_rp}. The calculation, carried out by solving the time-dependent Schr\"{o}dinger equation (\ref{diff1}) with $\langle \mathbf{J^{\ast}}^2 \rangle$ at each distance $r=r^{\ast}$, was based on the same molecular parameters as the above cw result. Marked in the figure is a particular value, $\tilde I$, of the laser intensity which is associated with a particular critical internuclear distance, $\tilde r^{\ast}$, as well as the ranges of the internuclear distance $r_p$ which do or do not lead to dissociation. We note that before reaching the asymptotic dependence $I \propto (r^{\ast})^3$ at $r^{\ast}>10$~\AA, the pulse intensity first decreases and passes through a minimum at about $r_{min}^{\ast}=5.8$ \AA, up to which the $I(r^{\ast})$ dependence is two-valued. Nevertheless, the short-distance region may be included in the calculation by simply changing the lower limit of integration in eq.~(\ref{DissProbGeneral}) from zero to some ``inner critical distance.'' However, since the time the molecule spends in the $r^{\ast}<r_{min}^{\ast}$ range is by a factor of about $2\times10^5$ times less than the time in spends at $r^{\ast}\ge r_{min}^{\ast}$, it is a fairly good approximation to neglect any effects that may arise at $r^{\ast}<r_{min}^{\ast}$. 

\begin{figure}
\includegraphics[width=8.5cm]{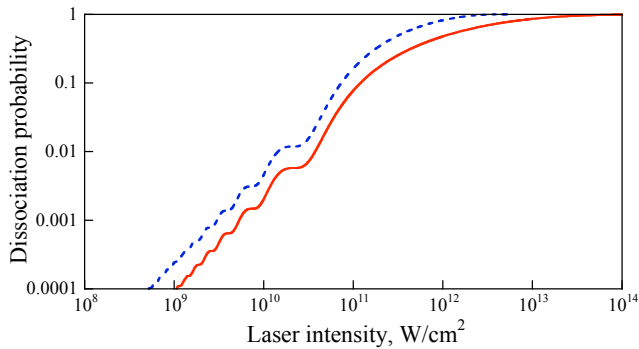}
\caption{\label{fig:Rb2_F_I} The dissociation probability of Rb$_2(^1\Sigma^+,v=123)$ as a function of the intensity of 50 ps nonresonant laser pulses. Calculation based on the ``exact" potential are shown by the solid red line, the analytic result by the dashed blue line. See text.}
\end{figure}

The vibrational probability density, Fig.~\ref{fig:Pots_WFs}, combined with the dependence of the laser intensity on the critical distance $I(r^{\ast})$, Fig. ~\ref{fig:Rb2_I_rp}, yields the dissociation probability, Eq.~(\ref{DissProbGeneral}), as a function of the laser intensity applied, $F(I)$. This is shown in Figure~\ref{fig:Rb2_F_I} by the full curve. Hence a measured dissociation curve  can be used, by reversing the procedure, to reconstruct the vibrational probability density and hence the potential that binds it.

The process of dissociation by shaking is amenable to an analytic treatment based on a near-threshold expansion of the wavefunction \cite{Harald}. The analytic wavefunction, obtained for the last bound state of the $^{85}$Rb$_2$ dimer, and the corresponding dissociation probability, Eq.~(\ref{DissProbGeneral}), are shown in Figs.~\ref{fig:Pots_WFs} and \ref{fig:Rb2_F_I}, respectively. The oscillations observed in the exact $F(I)$ are thus identified as due to the nodes of the vibrational probability density, and the position of the wavefunction's maximum as corresponding to the point of maximum slope, ``the edge,'' of $F(I)$.

We also assessed the dissociation probability of $^{39}$K$^{87}$Rb in its highest vibrational state, $v=99$. Since the molecule is polar, we invoked half-cycle pulses, with $\tau=50$~ps, to probe it. For a potential taken from ref.~\cite{Pashov07} and the body-fixed permanent dipole moment dependence on $r$ from ref.~\cite{Aymar05}, we found that the molecule can be shaken enough to dissociate  by laser intensities as low as $10^4$--$10^8$~W/cm$^2$ for $r \le 10$~\AA. As the permanent dipole moment plummets with increasing $r$, at higher internuclear separations the polarizability interaction prevails and there is no advantage in using the half-cycle pulses.  

We note that a non resonant cw-laser field can be used to control the scattering length and positions of magnetic Feshbach resonances in a simpler way than recently described by Bauer~\textit{et al.}~\cite{Bauer09}. Furthemore, the laser field of optical dipole traps may be able to dissociate some of the weakest-bound molecules or to reduce their apparent binding energy. However,  a 100~mW laser beam focused to a waist of 50~$\mu$m will not, as it imparts only a negligible angular momentum of $\langle \mathbf{J}^2 \rangle \approx 10^{-11}$ to a $^{85}$Rb$_2$ molecule. On the other hand, a cw field may be used to control the ``size" of a halo molecule, i.e., $\langle r_v \rangle \equiv \langle \phi_v(r)|r|\phi_v(r)\rangle$, by uplifting the highest vibrational state $v$. For instance, $^{85}$Rb$_2(v=123)$, which has a field-free mean radius $\langle r_v \rangle = 167$ \AA,~is expected to become a near-threshold state at an intensity of  $6.52 \times 10^8$ W/cm$^2$ (corresponding to $J=0.217$), with a mean radius of 1320 \AA. This represents an alternative to controlling molecular size by changing the magnetic field in the vicinity of a Feshbach resonance~\cite{KohlerRMP06}. Conversely, the nonresonant field can be used to provide a tunable barrier to molecule formation.

We are grateful to Gerard Meijer for discussions and support, and to Olivier Dulieu for discussions and for kindly providing the molecular polarizabilities used in this
work. We also thank to Francesca Ferlaino, Hanns-Christoph N\"agerl, Johannes Hecker-Denschlag, and Rudi Grimm for sharing with us their insights on Feshbach molecules.



\end{document}